\documentstyle[psfig,conf_iap,10pt]{article}
\begin{document}
\heading{%
%Begin Heading
%
New Territory: SZ Cluster Surveys\\
%
%End Heading
} 
\par\medskip\noindent
\author{%
%Begin Author names
James G. Bartlett$^{1}$, Alain Blanchard$^{1}$ \& Domingos Barbosa$^{2}$\\
%End Author names
}
\address{%
%First address
Observatoire de Strasbourg, 11 rue de l'Universit\'e, 67000 Strasbourg,
France.
}
\address{%
% Second Address
Astronomy Centre, University of Sussex, Falmer, Brighton BN1 9QJ, UK. 
}
%\address{%
% Third Address
% Here
%
%}

\begin{abstract}
	The potential  of the Sunyaev-Zel'dovich (SZ) effect
for cluster studies has long been appreciated, although
not yet fully exploited.  Recent technological advances
and improvements in observing strategies have changed this,
to the point where it is now possible to speak of this
subject at a meeting devoted to {\em surveys} in Cosmology.
We will discuss SZ surveys by distinguishing 
what may be called {\em pointed surveys}, dedicated to 
pre--selected clusters, from {\em blind surveys}, those
searching for clusters in blank fields.  Surveys of the 
former type already have significant numbers of clusters
with very good signal--to--noise images; surveys of the
second type are currently possible, but as yet not undertaken.
The discussion will focus on the kind of science that can
be done in this ``new territory''.  
\end{abstract}
\section{Introduction}
	Cluster studies based on the Sunyaev--Zel'dovich (SZ) effect
\cite{sz} have had a long and hard history of learning about 
systematic effects.  Benefiting from these early efforts and
thanks to new technologies (bolometers) and observing techniques
(interferometry), the field is now in rapid expansion.  At one
time the issue was simply one of detection; today, we speak of 
several tens of clusters with high signal--to--noise SZ {\em maps}.
The change has been revolutionary in its rapidity, and it has taken
place somewhat unbeknownst to those outside the field.  

	In this contribution, we will summarize what is happening 
by discussing the kind of astrophysics one can do with the SZ effect 
and the impact of current observations.  By far, the best and
most complete recent review of the subject is given by 
Birkinshaw \cite{birk}.  We have been inspired both by this
review and by the review given by Lange \cite{lange}.
We begin with a few general remarks concerning the SZ effect. 

\section{Properties of the SZ effect}

	There are some aspects of the SZ effect particularly worth emphasizing.
Recall that the effect is a distortion of the cosmic microwave
background (CMB) spectrum that may be expressed as $\delta i_\nu
= y(\vec{\theta}) j_\nu$, where $\delta i_\nu$ is the change in surface
brightness of the CMB at position $\vec{\theta}$ on the cluster face
relative to the unperturbed CMB spectrum.  The quantity $j_\nu$ 
gives the spectral shape of the distortion and is only
a function of frequency.  The magnitude is set by the
{\em Compton $y$-parameter}, an integral of the electron pressure
along the line--of--sight:
\begin{equation}
y \equiv \int dl\; \frac{kT_e}{m_ec^2} n_e \sigma_T
\end{equation}
Here, the electron temperature, mass and density are
$T_e$, $m_e$ and $n_e$, respectively, and $\sigma_T$ is
the Thompson cross section.  Integrating over the cluster
face, we find the {\em total} SZ flux density:
\begin{equation}
S_{sz} = \int d\Omega\; i_\nu(\vec{\theta}) = j_\nu D_a^{-2}(z)
	\int dV\; \frac{kT_e}{m_ec^2} n_e\sigma_T
\end{equation}
where $D_a$ is the angular--size distance.  There are two
important observations to make: 1/ $D_a = (1+z)^{-2} D_l$, i.e., 
the effect falls off with redshift much more slowly than
does, say, an X--ray flux, which is simply the well known
fact that the surface brightness of the effect is independent
of redshift for a fixed cluster; 2/ we may rewrite the expression
as $S_{sz} \sim f_{gas} M_{tot} <T_e>$ in terms of $f_{gas}$, the
hot gas mass fraction, $M_{tot}$, the total virial mass of the
cluster, and an average temperature defined by this expression,
$<T_e>\sim (1/N_e)\int dV\; n_e T_e$, where $N_e$ is the total
number of electrons in the cluster.  This is an important
aspect of the SZ effect, because the average temperature involved
is truly the {\em mean temperature} of the electrons in the
gas, as opposed, say, to the emission--weighted, X--ray
measured temperature.  The SZ--defined temperature is directly related
to the total energy of the electrons gained during 
collapse and {\em totally insensitive} to the spatial 
distribution of the hot gas in the cluster.  These two
points distinguish the SZ distortion from the X--ray flux
of a cluster and can be used to advantage in cluster
studies. 

\section{Pointed Surveys}

	As discussed by Birkinshaw \cite{birk}, there are three
techniques used to observe the SZ effect: traditional single--dish
observations (see, e.g., \cite{myers}), bolometers 
(see, e.g., \cite{suzie}) and interferometry
(see, e.g., \cite{bima} and \cite{rt}).  Interferometric 
observations now routinely produce SZ maps at more than 20 $\sigma$
at the center.  Bolometer observations have also detected the 
SZ effect in emission \cite{pronaos}.  
	
	Such observations can be used to deduce cluster baryon
fractions \cite{myers}, constrain (eventually detect) cluster 
peculiar velocities \cite{nabila}, \cite{vpec} and 
to trace total cluster mass profiles.  In the former and latter cases, 
the density insensitivity of the SZ effect is a pertinent advantage
over X--ray studies (SZ varies linearly with $n_e$).  
Combined X--ray and SZ observations 
lead to distance estimates and a Hubble diagram.  Figure 1 shows
such a diagram from data we gleaned from the literature 
(the diagram was inspired by Birkinshaw \cite{birk}, although
the data is not exactly the same).
\begin{figure}
\centerline{\vbox{
\psfig{figure=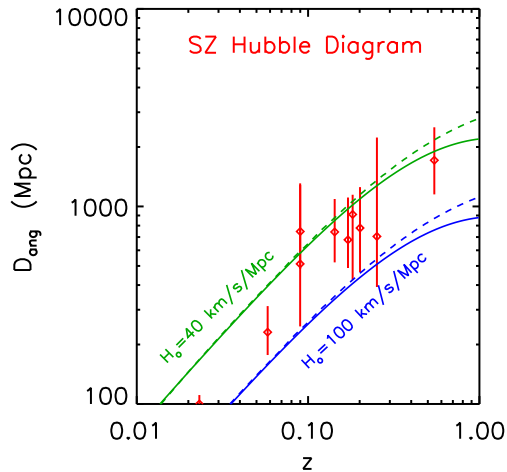,height=7.cm}
}}
\caption[]{Hubble diagram with SZ observations.  See also 
\cite{birk}
}
\end{figure}

\section{Blind Surveys}

	Here we refer to purely SZ surveys where clusters 
are selected from ``blank'' sky soley on the basis of their
SZ flux density.  Such surveys are just feasible today.
Surveys of this kind should be viewed as totally analogous
to X--ray surveys (e.g., \cite{bs}): one can study SZ luminosity functions and
their evolution (closely related to the amount and evolution
of the intracluster medium), the cluster contribution to 
the mean background (mean distortion of the CMB spectrum) and
to the background anisotropies, and redshift 
distributions.  As discussed in \cite{sz1}, 
\cite{sz2}, the
latter can be used to constrain the density parameter, $\Omega$.
Particularly enticing are the reports of SZ decrements with
no optical or X--ray counterparts \cite{vla}, \cite{rthole}.
If truly due to clusters, these must be at very large redshifts
($z>1$), a fact difficult to account for in a critical Universe
\cite{sz2} (see Figure 2).

\begin{figure}
\centerline{\vbox{
\psfig{figure=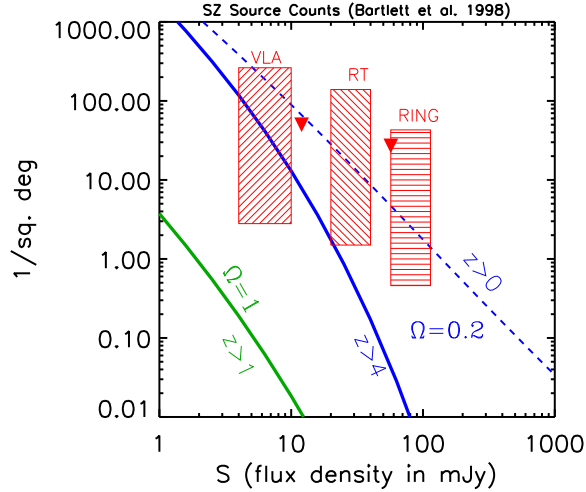,height=7.cm}
}}
\caption[]{SZ source counts confronted with the two possible
SZ clusters found by the VLA \cite{vla} and Ryle Telescope
\cite{rthole}.  If the decrements are truly clusters, they
must be at $z>1$ to satisfy the X--ray flux upper limits;
this would be difficult to explain in an $\Omega=1$ Universe.
This figure was taken from \cite{sz2}
}
\end{figure}

\section{Conclusion}

	We are now in a position to reap the potential of SZ observations.
Catalogs of pre--selected clusters observed in SZ (pointed surveys) 
already count several tens of entries, permitting for the first time
statistical studies.  True blank field surveys will soon be undertaken
over small patches of sky with interferometers and, eventually, with
large bolometer arrays \cite{lange}; over the longer term, 
one anticipates a SZ--selected catalog of more than 10,000 sources
from the Planck Surveyor \cite{nabila}, a remarkable opportunity.

\acknowledgements{Special thanks go to A. Cooray for answering
many questions, helping much, and for providing some beautiful 
figures.}

\vfill
\end{document}